\documentstyle[11pt, aasms4, epsf]{article}

\def\aa#1#2#3#4#5{\bibitem[#1]{#2}#3, {\it A\&A}, {\bf#4}, #5.}

\def\apj#1#2#3#4#5{\bibitem[#1]{#2}#3, {\it Ap. J.}, {\bf#4}, #5.}
\def\apjlett#1#2#3#4#5{\bibitem[#1]{#2}#3, {\it Ap. J. (Letters)}, {\bf #4},
#5.}

\def\icarus#1#2#3#4#5{\bibitem[#1]{#2}#3, {\it Icarus}, {\bf #4}, #5.}

\def\nature#1#2#3#4#5{\bibitem[#1]{#2}#3, {\it Nature}, {\bf #4}, #5.}

\def\pasp#1#2#3#4#5{\bibitem[#1]{#2}#3, {\it PASP}, {\bf#4}, #5.}

\def\science#1#2#3#4#5{\bibitem[#1]{#2}#3, {\it Science}, {\bf #4}, #5.}

\begin{document}

\title{Sensitivity of the Radial Velocity Technique in Detecting 
Outer Planets}

\author{J. A.  Eisner \& S. R. Kulkarni}
\affil{Palomar Observatory 105-24, California Institute of
Technology, Pasadena, CA 91125}

\begin{abstract}
The technique of radial velocity (RV) has produced spectacular
discoveries of short-period Jovian mass objects around a fraction 
(5 to 10\%) of nearby G stars.  Although we expect Jovian planets
to be located in long-period orbits of decades or longer (if our 
solar system is any guide), detecting such planets
with RV technique is difficult due to smaller
velocity amplitudes and the limited temporal baseline (5--10 yr) of
current searches relative to the expected orbital periods. 
In this paper, we develop an analytical understanding
of the sensitivity of RV technique in the regime where the the 
orbital period is larger than the total baseline of the survey.
Moreover, we focus on the importance of the orbital phase
in this ``long-period'' regime, and develop a Least Squares detection
technique based on the amplitude and phase of the fitted signal.  

To illustrate the benefits of this amplitude-phase analysis, we
compare it to existing techniques.  
Previous authors ({e.g.} \cite{NA98})
have explored the sensitivity of an amplitude-only analysis
using Monte Carlo simulations.  Others have supplemented this 
by using the slope of the linear component of the fitted sinusoid
in addition ({e.g.} \cite{WALKER+95}; \cite{CMB99}).  
In this paper, we illustrate the benefits of
Least Squares over periodogram analysis, and demonstrate the 
superiority of an amplitude-phase technique over previous analyses.

\end{abstract}

\vfill\eject

\section{Introduction}
\label{sec:intro}

Radial velocity surveys of nearby stars have been employed in the
search for extra-solar planets for nearly two decades.  An important
element of these surveys is the sensitivity; {i.e.} what is the
minimum-mass planet that can be detected at a given orbital period?
This issue boils down to a question about how well one can detect
periodic signals in a finitely sampled data set.  

There are two main regimes of analysis:  the ``short-period'' regime,
characterized by $\tau \ll T_0$ and the ``long-period'' regime with
$\tau \gg T_0$.  Here $\tau$ is the orbital period and $T_0$ is the
duration of the survey.  The planetary objects found to date through
the RV technique are short period objects, with periods less than a
year or so ({e.g.} \cite{MQ95}; \cite{MB96}; \cite{FISCHER+98}).  The
sensitivity of RV searches is very well understood in this short
orbital period regime and compact analytical formulae exist
(\cite{LOMB76}; \cite{SCARGLE82}; \cite{HB86}; \cite{NA98};
Cumming, Marcy, \& Butler 1999).

According to current theoretical prejudice, however, one expects giant
planets to be primarily formed in the colder regions of the
proto-planetary nebula, and thus we expect such objects to possess
periods in the range of many years to centuries (\cite{BOSS95}).
Unfortunately, it is well known that the sensitivity of RV technique is
considerably worse for planets with large orbital periods.  Earlier
analyses in the long-period regime have primarily concentrated on
estimating the sensitivity through simulations (e.g. \cite{NA98},
\cite{CMB99}).

This paper has two goals: to provide analytical insight
into the sensitivity of the RV technique  in the long-period regime,
and to address the issues of
detection and detection efficiency. We note that almost all of the
previous work has concentrated on setting upper limits and not
addressed the issue of detection. It is also worth noting that much of 
the previous work has been based on the periodogram method
({e.g.} \cite{SCARGLE82}; \cite{HB86}; \cite{WALKER+95}; \cite{CMB99}).  
Following,
Nelson \&\ Angel (1998) \nocite{NA98}, we argue at various points in the
paper why a Least Squares approach is preferable to a periodogram
approach. Essentially the  Least Squares approach, in contrast to the
traditional periodogram approach, offers the most general method and
needs no modifications in the long-period regime or in the sparse-data
regime.

The plan of the paper is as follows. We summarize the Least Squares
approach and derive the basic equations in \S\ref{sec:basics}.  In
\S\ref{sec:typeI-errors} we provide analytical estimates for obtaining a
detection in the absence of any signal in the short and long-period regimes. 
In \S\ref{sec:typeII-errors} we carry out simulations and obtain estimates
for minimum detectable signals in the presence of noise.
We conclude in \S\ref{sec:conclusions}.

\section{Radial Velocity Technique: Basic Equations}
\label{sec:basics}

For simplicity, we will assume circular orbits throughout this
discussion.  A planet in a circular
orbit undergoes acceleration, and because the linear momentum of the system
must be conserved, the star undergoes a reflex acceleration.  It is
this acceleration that directly informs us of the presence of the
companion. However, the observable is the velocity:
\begin{equation}
	v(t) = A \sin(2\pi t/\tau + \phi) + \gamma
\label{eq:BasicRV}
\end{equation}
where the amplitude, $A$ is given by 
\begin{equation}
A = \left(\frac{2\pi G M}{\tau}\right)^{\frac{1}{3}} \: \frac{M_p \sin i}
{M_{\ast}}.
\label{eq:Amplitude}
\end{equation}
Here, 
$\gamma$ is the radial velocity of the planetary system,
$\phi$ is the orbital phase,
$\tau$ is the orbital period, $M_p$ is the planet mass, $M_{\ast}$
is the stellar mass, $M=M_p+M_{\ast} \approx M_{\ast}$ (as $M_p \ll
M_{\ast}$), and $i$ is the inclination of the orbit with respect to the
plane of the sky. 

The sensitivity of an RV experiment is essentially defined as the 
minimum mass planetary companion that can be detected at a 
given period.
For a given measurement of $A$, we invert Equation 
\ref{eq:Amplitude} to obtain a relation between $M_p$ and $\tau$:
\begin{equation}
M_p \sin i= A
\left(\frac{M_{\ast}^2}{2\pi G}\right)^{\frac{1}{3}} 
\: \tau^{\frac{1}{3}} .
\label{eq:Mpsini}
\end{equation}

We see that $M_p \propto \tau^{\frac{1}{3}}$, which
means that our sensitivity decreases as we look for longer-period orbits.
Thus, it is not surprising that the first detection of
an extra-solar planet was a planet with very short orbital period
(\cite{MQ95}). There is an additional effect which makes
it difficult to identify planets with long periods. With only
a baseline of a few years, the RV surveys 
can only observe
some fraction of $A$ for orbital periods longer than the observing 
baseline. 
In this (long-period) regime, the sensitivity depends critically
on the orbital phase.
The most sensitive mass estimates are obtained when the
acceleration in the radial direction attains the extreme value.
In contrast, when the radial acceleration is close to zero, 
the lack of curvature in the orbit makes the signal
difficult to distinguish from the unknown systemic velocity of
the system.
These two effects combine to explain why all the  planetary 
detections  to date have  orbital periods of less than 3 years 
(\cite{MQ95}; \cite{MB96}; \cite{BM96}; \cite{BUTLER+97};  
\cite{NOYES+97}; \cite{CHBM97};  \cite{MARCY+98a}; \cite{FISCHER+98}; 
\cite{MAYOR+98}; \cite{MARCY+99}; \cite{QUELOZ+99}).

\subsection{Least Squares Fitting of Sinusoids}
\label{sec:LSq}

The basic RV analysis consists of fitting the
observations to the model specified in Equation~\ref{eq:BasicRV} and
then inferring the mass of the planet[s] from the fitted
values through Equation~\ref{eq:Mpsini}. As noted by several authors
(e.g. \cite{SCARGLE82}; \cite{HB86}; \cite{NA98}) the most optimal 
fitting method is obtained by using the technique of Least Squares.
To enable the use of linear Least Squares fitting techniques
we proceed by deriving a linear model equation from 
Equation~\ref{eq:BasicRV}:
\begin{equation}
	v(t) = 
v_c \cos(\omega t_i) + v_s \sin(\omega t_i) + \gamma;
\label{eq:LSQmodel}
\end{equation}
here $v_c=A\sin{\phi}$, $v_s=A\cos{\phi}$ and
$\omega=2\pi/\tau$.  

There appears to be considerable discussion and debate about the
$\gamma$ term in the literature (e.g. \cite{WALKER+95}; \cite{CMB99}).  
The origin of
this debate can be traced to two issues. First, it is an assumption of
periodogram analyses that the signal is a pure sinusoidal wave (i.e.
$\gamma=0$). Indeed, the classic periodogram problem is the sine wave
(with no offset) buried in zero-mean noise.  The second issue is that
astronomers are rarely interested in determining the precise radial
velocity of  a star and its planetary system--i.e. the systemic motion.
For this reason, $\gamma$ is usually seen as a ``nuisance'' parameter.
In some periodogram analyses, 
$\gamma$ is first determined from the mean of the
data and in others $\gamma$ is determined in conjunction with $v_c$ and
$v_s$. Cumming et al. (1999) refer to these as ``fixed'' and
``floating'' mean methods, respectively while Walker et al. (1995) call
them ``correlated'' and ``uncorrelated''.
However, our view is that such methods for including $\gamma$
(and the additional extension
of including a constant acceleration term; see 
\S\ref{sec:linear}) 
only make the periodogram closer to the Least Squares method.  Thus,
on philosophical grounds of generality 
we prefer the Least Squares approach.  We justify this choice
in greater detail below.

Essentially the basic fact is that $\gamma$ is {\it needed} to
represent the physical model correctly. $\gamma$ may be an
uninteresting parameter, but it is as unknown as $v_c$, $v_s$ and $\tau$.
$\gamma$ can be dropped from Equation~\ref{eq:BasicRV} only if it can
be demonstrated that it is not covariant with the remaining three
parameters, $v_c$, $v_s$ and $\tau$. As shown below, this is not the
case and thus one must solve for all the parameters simultaneously.  In
astrometry, the problem is considerably worse with the position and
proper motion being covariant with the parameters of a potential
planetary orbit. As with the RV case, one must solve for all unknown
parameters simultaneously rather than sequentially, and the periodogram
analyses would then include ``floating means''and ``first derivatives''.

Only in the short-period regime can $\gamma$  potentially get
decoupled from the other parameters. For this to happen, we need dense
sampling over a number of cycles.  In the long-period regime, as noted
below, the cross-talk never disappears and one must solve for $\gamma$,
regardless of the density of sampling.  The covariance of $\gamma$ with
orbital parameters is easily seen here since  in this regime, we only
measure a portion of the orbit and in this limit, the orbit can be
approximated by a linear term (constant velocity) and curvature
(acceleration). The first term is covariant with the systemic motion.
In the astrometry context, Black \&\ Scargle (1982) were the first to
recognize the consequences of this covariance.

The parameters of our data set are as follows: the duration of the 
survey is $T_0$, $v'(t_j)$ is the
measured RV at epoch $t_j$, and $n_0$ is the number of measured
epochs. With no loss of generality, we let
our time go from $t=-T_0/2$ to $t=T_0/2$. This device allows us to
simplify the normal equations (see below).
To find the three unknowns, $v_c$, $v_s$ and $\gamma$, we minimize
\begin{equation}
X^2 = \sum_{i=0}^{n_0-1} [v'(t_i) - v_c \cos(\omega t_i)
- v_s \sin(\omega t_i) - \gamma]^2
\label{eq:LSQEq}
\end{equation}
with respect to $v_c$, $v_s$, and $\gamma$.
This yields a matrix
equation for the three unknowns:
\begin{displaymath}
\left(\matrix{\sum_{i=0}^{n_0-1} \cos^2(\omega t_i) & 
\sum_{i=0}^{n_0-1} \cos(\omega t_i) \sin(\omega t_i) &
\sum_{i=0}^{n_0-1} \cos(\omega t_i) \cr 
\sum_{i=0}^{n_0-1} \cos(\omega t_i) \sin(\omega t_i) &
\sum_{i=0}^{n_0-1} \sin^2(\omega t_i) &
\sum_{i=0}^{n_0-1} \sin(\omega t_i) \cr
\sum_{i=0}^{n_0-1} \cos(\omega t_i) & 
\sum_{i=0}^{n_0-1} \sin(\omega t_i) & n_0} \right) 
\times \left(\matrix{v_c \cr v_s \cr \gamma} \right)
\end{displaymath}
\begin{equation}
= \left(\matrix{\sum_{i=0}^{n_0-1} v'(t_i) \cos(\omega t_i) \cr
\sum_{i=0}^{n_0-1} v'(t_i) \sin(\omega t_i) \cr
\sum_{i=0}^{n_0-1} v'(t_i)} \right) .
\label{eq:NormalEqs}
\end{equation}

In the short-period regime where $\tau\ll T_0$, we observe many cycles
and thus, under most reasonable sampling schemes, the 
sinusoidal summations, $\sin(\omega t)$, $\cos(\omega t)$, and 
$\sin(\omega t) \times \cos(\omega t)$ will average to zero,
while $\sin^2(\omega t)$ and $\cos^2(\omega t)$ 
will average to 1/2.  Thus, in the short-period
regime the matrix in 
Equation~\ref{eq:NormalEqs} becomes diagonal, and the fit
parameters are given by

\begin{equation}
v_c = \frac{2}{n_0} \sum_{i=0}^{n_0-1} v'(t_i) \cos(\omega t_i),
\label{eq:vc-short}
\end{equation}

\begin{equation}
v_s =\frac{2}{n_0}  \sum_{i=0}^{n_0-1} v'(t_i) \sin (\omega t_i),
\label{eq:vs-short}
\end{equation}

\begin{equation}
\gamma = \frac{1}{n_0} \sum_{i=0}^{n_0-1} v'(t_i).
\label{eq:gamma-short}
\end{equation}

In the long-period regime, $\tau > T_0$,  
the matrix is not diagonal since
most of the terms do not average to zero. However, by design
(and assuming a reasonable sampling scheme), the sinusoidal
summations involving only one power of  $\sin(\omega t$) 
will average to zero yielding the following normal equations:
\begin{equation}
v_c \sum_{i=0}^{n_0-1} \cos^2(\omega t_i) + 
\gamma \sum_{i=0}^{n_0-1} \cos(\omega t_i) = 
\sum_{i=0}^{n_0-1} v'(t_i) \cos(\omega t_i)
\label{eq:normal-long1}
\end{equation}
\begin{equation}
v_s \sum_{i=0}^{n_0-1} \sin^2(\omega t_i) =
\sum_{i=0}^{n_0-1} v'(t_i) \sin(\omega t_i)
\label{eq:normal-long2}
\end{equation}
\begin{equation}
v_c \sum_{i=0}^{n_0-1} \cos(\omega t_i) +
\gamma n_0 = \sum_{i=0}^{n_0-1} v'(t_i).
\label{eq:normal-long3}
\end{equation}

\section{Type I Errors}
\label{sec:typeI-errors}

Type I errors describe the probability that a high amplitude will be
obtained even when no signal is present in the data.  We assess the
probability of type I errors, assuming that our data set consists of
Gaussian noise with a mean of zero and a standard
deviation of $\sigma_0$.

\noindent{\bf Short-Period Regime.} 
From Equations~\ref{eq:vc-short}--\ref{eq:gamma-short} we note that
for most reasonable sampling schemes,
$v_c$, $v_s$ and $\gamma$ are simply sums of Gaussian variables
and thus from the Gaussian addition theorem, all three derived
parameters are also Gaussian variables.
Specifically, $v_c$ and $v_s$ obey a Gaussian distribution
with a mean of zero and a standard deviation of
\begin{equation}
\sigma = \sqrt{\frac{2}{n_0}} \sigma_0,
\label{eq:Sigma}
\end{equation}
where $n_0$ is the number of measurements.
Denoting by $V_{1s}$ the value of $\vert v_c \vert$ (or
$\vert v_s \vert$) that is exceeded in 1\% of cases, we
note that
\begin{equation}
V_{1s}=2.61\sigma = \frac{3.69}{\sqrt{n_0}}\: \sigma_0.
\label{eq:V1s}
\end{equation}
From Equation~\ref{eq:gamma-short} we see that $\gamma$ follows
a Gaussian distribution with zero mean and a variance of $\sigma^2/n_0$.
Denoting $\Gamma=\vert\gamma\vert$ we note that the 
99$^{\rm th}$ percentile value
of $\Gamma$ is $\Gamma_{1s}=2.61\sigma_0/\sqrt{n_0}$. Thus 
$\Gamma_{1s} = 0.71V_{1s}$.

Because we are interested in the fitted amplitude, we will now
examine the combined statistics of $v_c$ and $v_s$.
The probability density function for 
$v_c$ and $v_s$ is
\begin{equation}
\rho(v_c,v_s) dv_c dv_s = \frac{1}{2 \pi \sigma^2} \: e^{-v_c^2/2\sigma^2}
\: e^{-v_s^2/2\sigma^2} \: dv_c \: dv_s .
\label{eq:P(vc,vs)}
\end{equation}
Denoting by $K$ the square of the inferred amplitude,
$K=v_c^2+v_s^2$, we note that the probability density function
of $K$ is an exponential and find the probability to be
\begin{equation}
P(K<K_0) = 1 - e^{-K_0/2\sigma^2}.
\label{eq:P(K)}
\end{equation}
Thus, the squared velocity amplitude 
that is exceeded by pure fluctuations
in only 1\% of the cases, is
\begin{equation}
K_{1s} = 2\sigma^2 \ln\left(\frac{1}{1-0.99}\right) 
= \frac{18.42 \: \sigma_0^2}{n_0}.
\label{eq:K1s}
\end{equation}

\noindent{\bf Long-Period Regime.}
In this regime, the expressions for $v_c$, $v_s$ and $\gamma$ are
not as simple as those in the short-period regime, and so we resort
to simulations. To this end,
we create a synthetic
data set consisting of Gaussian noise with $\sigma_0 = 3$ m s$^{-1}$,
which is the best accuracy thus far obtained for radial velocity
measurements of this type (\cite{BUTLER+96}).  We sample the synthetic data
at one month intervals for $T_0 = 12$ years.
We explore fitted periods from 5 years to 100 years, choosing the
interval between sampled periods so as to result in a 1 radian
decrease in the total number of orbital cycles over the the length of
the 
observations,
\begin{equation}
\Delta \tau = \frac{\tau^2}{2\pi T_0}.
\label{eq:DTau}
\end{equation}
Thus the sequence of the periods which we consider is 
60, 64, 69, 74, 80, 87, 95, 105, 117, 132, 152, 177,
212, 261, 337, 463, 699, and 1239 months.

\begin{figure}[t]
\vspace{5.0 in}
\includegraphics{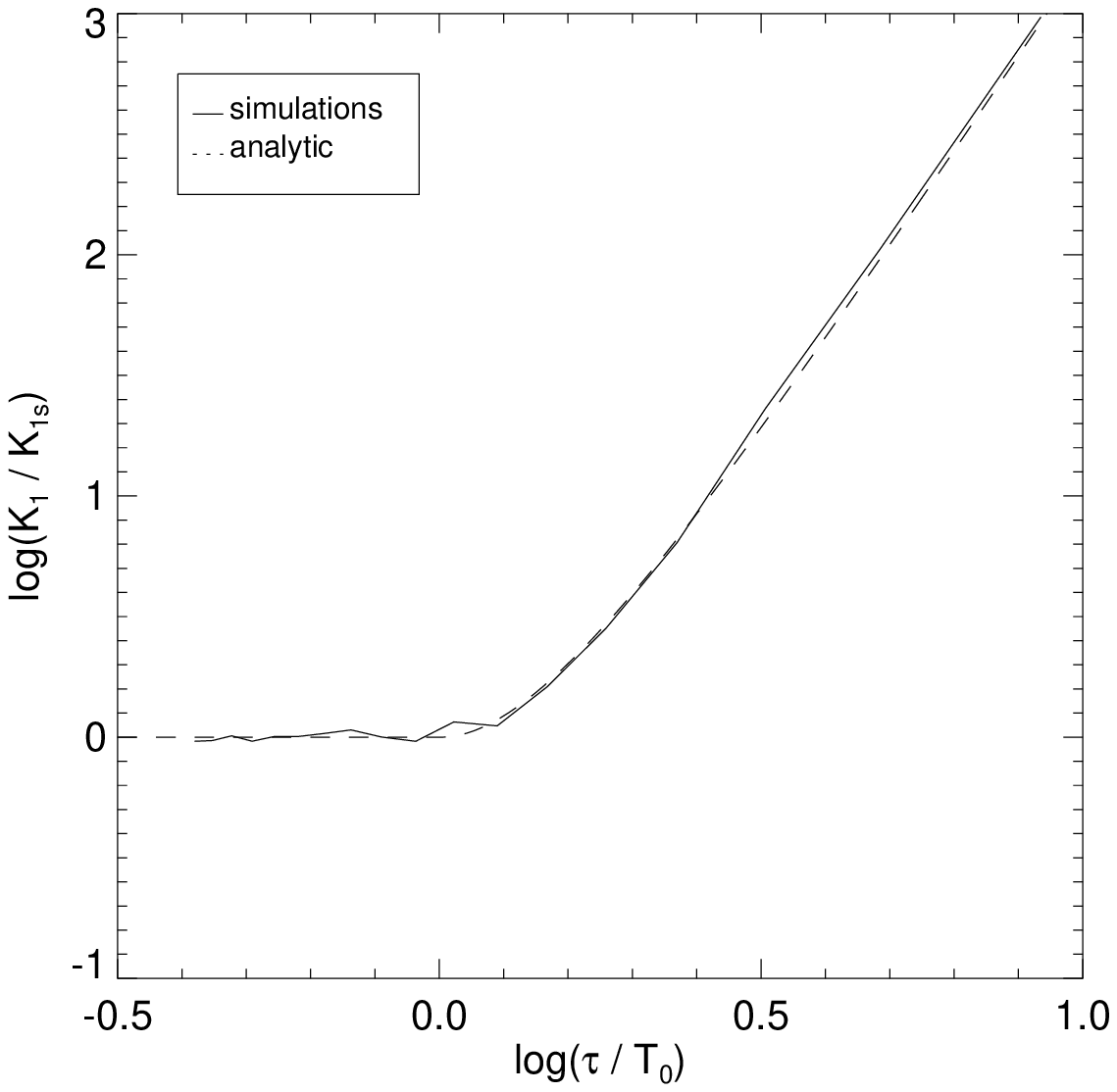}
\caption{A plot of $\log(K_1 / K_{1s})$ versus $\log(\tau / T_0)$ in 
the long-period regime of the radial velocity technique (solid line).  
$K_1$ is the 99th percentile of $K = v_c^2 + v_s^2$.
Here, we simulated $N=1000$ data sets with
Gaussian noise of zero mean and $\sigma_0 = {\rm 3 \: m \: s^{-1}}$.
The data were sampled at one month intervals for $T_0=144$ months.
The dashed line shows the analytical $K_{1}$ predicted by Equation
\ref{eq:K1long}.
\label{fig:K1}}
\end{figure}

For each period, $\tau$, we simulate $N = 1000$ data sets and carry out
the Least Squares fit. We set $K_1$ equal to the 10$^{\rm th}$ highest
$K$ that arises.  Clearly, 99\% of the fitted $K$'s will lie below
this value.  A plot of $K_{1}$ versus $\tau$ is shown in Figure 
\ref{fig:K1}.  As Figure \ref{fig:K1} indicates, our simulations 
are in excellent agreement with the expected
value of $K_{1s}$ in the short period regime (Equation~\ref{eq:K1s}).

Thus far, our analysis has followed that of Nelson \& Angel (1998), but at
this point the two analyses diverge.  Nelson \& Angel (1998) assume that
$K_1$ in the long-period regime is  a power law and from their
simulations find $K_1 \propto \tau^{3.64}$. The value of this
exponent has no natural
explanation and indeed it was this unusual value that
motivated the present analysis.

The principal goal 
of our analysis was to understand the behavior of $K_1$ in the
long-period regime and to derive an analytical expression for $K_1$
that is valid for all fitted periods.
We begin by examining the covariance matrix of $v_c$, $v_s$, and $\gamma$.
At short periods, we find these three fitted parameters are 
uncorrelated; this is expected from 
Equations \ref{eq:vc-short}--\ref{eq:gamma-short}.

\begin{figure}
\vspace{7.5 in}
\includegraphics{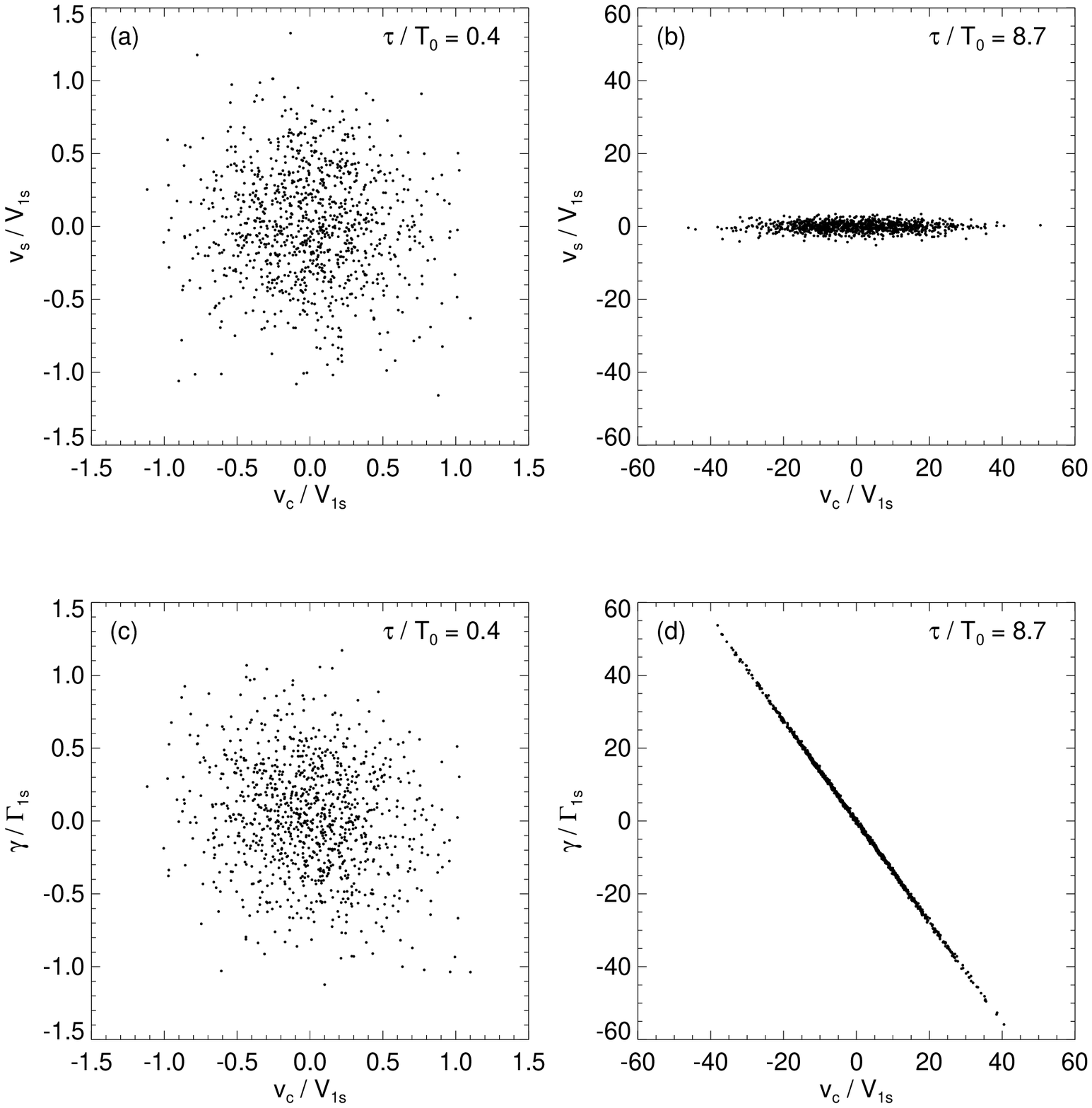}
\caption{Plots of $v_c$ versus $v_s$ and 
$v_c$ versus $\gamma$ in the short and long-period regimes.
In both cases, the duration of the RV monitoring is $T_0=12$ yr.
In the short-period regime, $\tau \ll T_0$, $v_c$, $v_s$,
 and $\gamma$ are uncorrelated (a),(c). In the long-period regime,
strong correlations are seen. In particular, we see that
orbital phases of approximately $\pm 90^{\circ}$ are preferred (b)
and that 
$v_c$ and $\gamma$ are strongly anti-correlated (d). 
\label{fig:corrls}}
\end{figure}

However, the situation is quite different in the long-period regime as
can be seen from the plots in
Figure \ref{fig:corrls}.  
From this figure we conclude the following:
\newline 1.  $v_c$ and $v_s$ are correlated. 
\newline 2.  $v_c$ and $\gamma$ are anti-correlated. 
\newline 3.  $v_s$ and $\gamma$ are uncorrelated. \newline
\noindent
The corollary to this result is that phase becomes important in the
long-period regime. Indeed, as can be seen from Figure~\ref{fig:corrls},
$\phi \sim \pm 90^\circ$ is preferred.  Nelson \& Angel (1998) and 
Cumming et al. (1999) \nocite{CMB99}
recognized that the phase would become non-random in the long period
regime. However, neither they nor others have explored the full
implications of this fact. Below we look into this issue in more
detail.

We now provide a simple (physical) explanation of the results
displayed in Figure~\ref{fig:corrls}.
When $\tau \ll T_0$, sinusoids with random phase can be fit to 
the Gaussian data.  Clearly, the amplitude of these sinusoids
cannot be significantly bigger than the vertical scale of the data
(which is approximately $\sigma_0 = {\rm 3 \: m \: s^{-1}}$).
However, when $\tau > T_0$, this is no longer necessarily 
the case.  The maximum value of $K$ is obtained when we fit
a cosine to random data since a cosine is flat around $t=0$;
we remind the reader of the choice of our
time baseline, $[-T_0/2, T_0/2]$. Recall that
for small $t$, $\cos t \simeq 1 - t^2/2$.
Thus, the size of the fitted cosine is limited by how much it 
deviates from a constant in the range from $t=[0,T_0/2]$.  This
deviation is given by
\begin{equation}
1-\cos\left(\frac{\pi T_0}{\tau}\right).
\label{eq:cos_dev}
\end{equation}

The amplitude of this cosine is then chosen so that this
deviation from a constant will be roughly equal to the vertical
scale of the noise ($\sigma_0$).  So, Equation~\ref{eq:cos_dev} 
tells us the fraction of the total fitted amplitude, and
thus the actual fitted amplitude is
\begin{equation}
V_{c_1} = \frac{2 V_{1s}}{(1-\cos\left[\frac{\pi T_0}{\tau}\right])};
{\rm \:\:\:\: for \:}\tau \ge T_0
\label{eq:VcLong}
\end{equation}
here, $V_{c_1}$ is the value of $\vert v_c\vert$ which will be exceeded in
1\% of Least Squares fits to Gaussian noise,
$V_{1s}$ is the corresponding $V_{c_1}$ in the short-period
regime (Equation \ref{eq:V1s}), 
and 2 is a normalization factor (since the above deviation was
peak-to-peak).

\begin{figure}[t]
\vspace{5.5 in}
\includegraphics{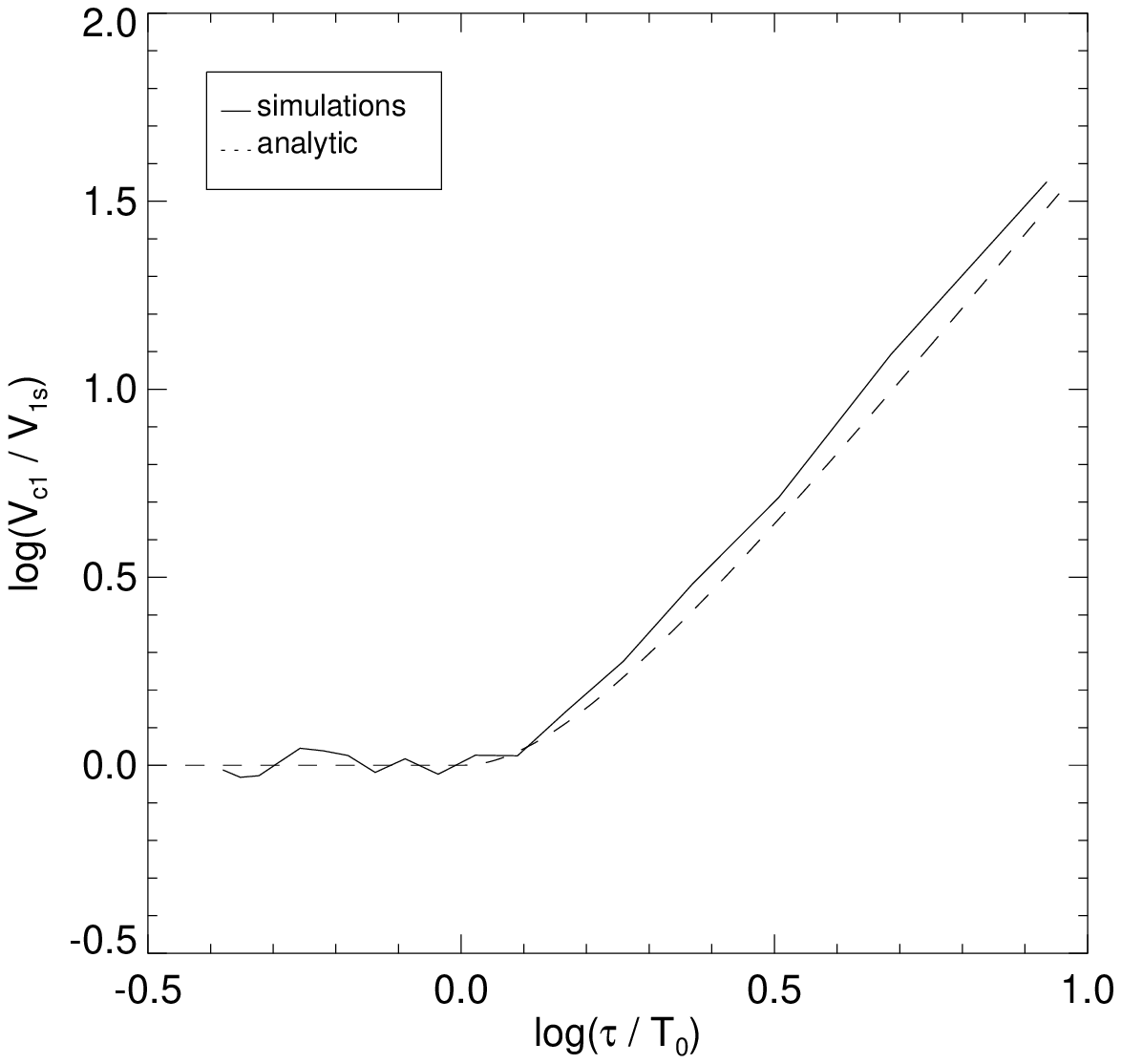}
\caption{A plot of $\log(V_{c_1}/V_{1s})$ versus
$\log(\tau/T_0)$.  $V_{c_1}$ is the value of $\vert v_c\vert$ that is
exceeded in 1\% of the simulations and $V_{1s}$ is a normalizing
factor (see Equation~\ref{eq:V1s}).  See the caption of Figure
\ref{fig:K1} for details of the simulations.
The solid line shows the behavior of
the simulated data, and the dashed line represents the
analytical expression from Equation \ref{eq:VcLong}. 
\label{fig:VC}}
\end{figure}

\begin{figure}[t]
\vspace{5.5 in}
\includegraphics{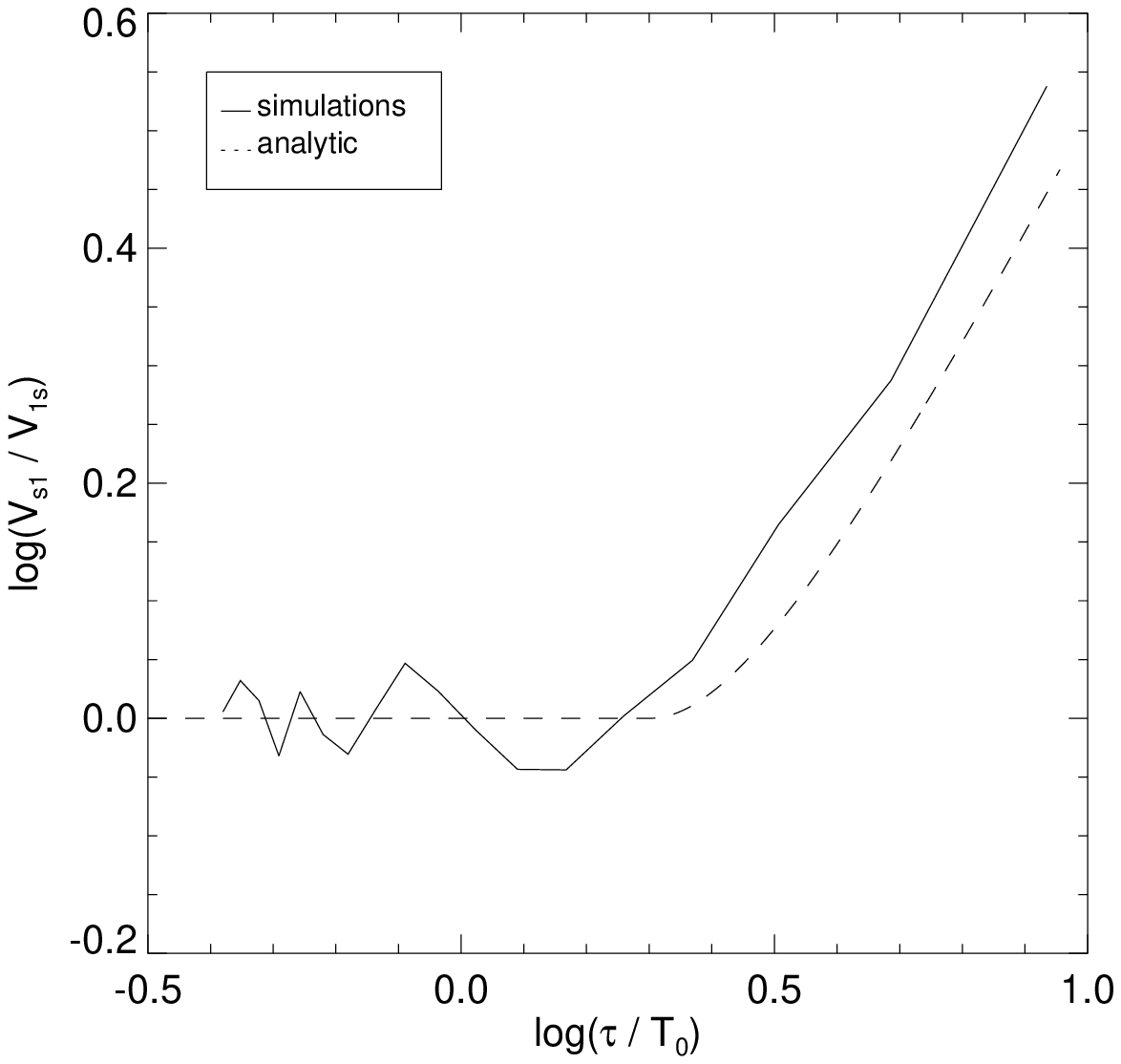}
\caption{
A plot of $\log(V_{s_1}/V_{1s})$ versus
$\log(\tau/T_0)$. $V_{s_1}$ is the value of $\vert v_s \vert$ that is
exceeded in 1\% of the simulations. 
See caption to Figure~\ref{fig:K1} for details of the simulations.  The
solid line shows the behavior of the simulated data, and the dashed
line represents the analytic expression from Equation \ref{eq:VsLong}.
Bearing in mind that the vertical scale of this figure is quite fine, 
we note that the analytical expression is a good fit to
the simulated data.  
\label{fig:VS}} 
\end{figure}

Now, we consider the behavior of $v_s$ in the long-period regime.
Here, there is a slight difference: because the full amplitude of 
a sine centered around $t=0$ can be observed in the interval
$[-\tau/4,\tau/4]$, the ``long-period regime for $v_s$'' does
not actually begin until $\tau > 2T_0$.  However, once we
are in the long-period regime, our analysis is much the same.
In particular, we find the amplitude of a sine wave such that
in the interval from $t=[-T_0/2,T_0/2]$, the sine wave does
not exceed the vertical scale of the noise.  Following the
same lines as the analysis for $v_c$, it is clear that this
amplitude must be
\begin{equation}
V_{s_1} = \frac{V_{1s}}{\sin\left(\frac{\pi T_0}{\tau}\right)};
{\rm \:\:\:\: for \:}\tau > 2T_0
\label{eq:VsLong}
\end{equation}
This function is plotted with the simulated data in Figure
\ref{fig:VS}. Not surprisingly, $V_{s_1}$ is considerably smaller
than $V_{c_1}$.

Understanding the behavior of $\gamma$ is not difficult in light
of our understanding of $v_c$. Since the mean value of the
simulated data is zero, our fitted function must also, in the mean,
be zero. The anti-correlation between $\gamma$ and $v_c$ in 
Figure~\ref{fig:corrls} confirms this expectation.
From the above discussion, we know that the fitted signal is primarily
a cosine with an amplitude given by Equation~\ref{eq:VcLong}.
We must choose a $\gamma$ that translates the
fitted cosine such that the result is ``centered'' 
around $y=0$.  This translation is accomplished by subtracting
the fitted amplitude $A$ (which produces a cosine wave whose maximum
value is zero) and then adding back $\sigma_0/2$ in order to
properly center the fitted signal.  Thus, we expect
\begin{equation}
\Gamma_1 = \Gamma_{1s} \left\{ \frac{3}{1-\cos\left[\frac{\pi T_0}{\tau}\right]} 
-\frac{1}{2} \right\}.
\label{eq:GammaLong}
\end{equation}
Note that we expect no correlation between $v_s$ and $\gamma$,
because $v_s$ is already centered around zero, and requires no
translation (or physically, a sinusoidal orbit with $\phi=0^{\circ}$
is non-degenerate with constant velocity motion).

\begin{figure}[t]
\vspace{5.5 in}
\includegraphics{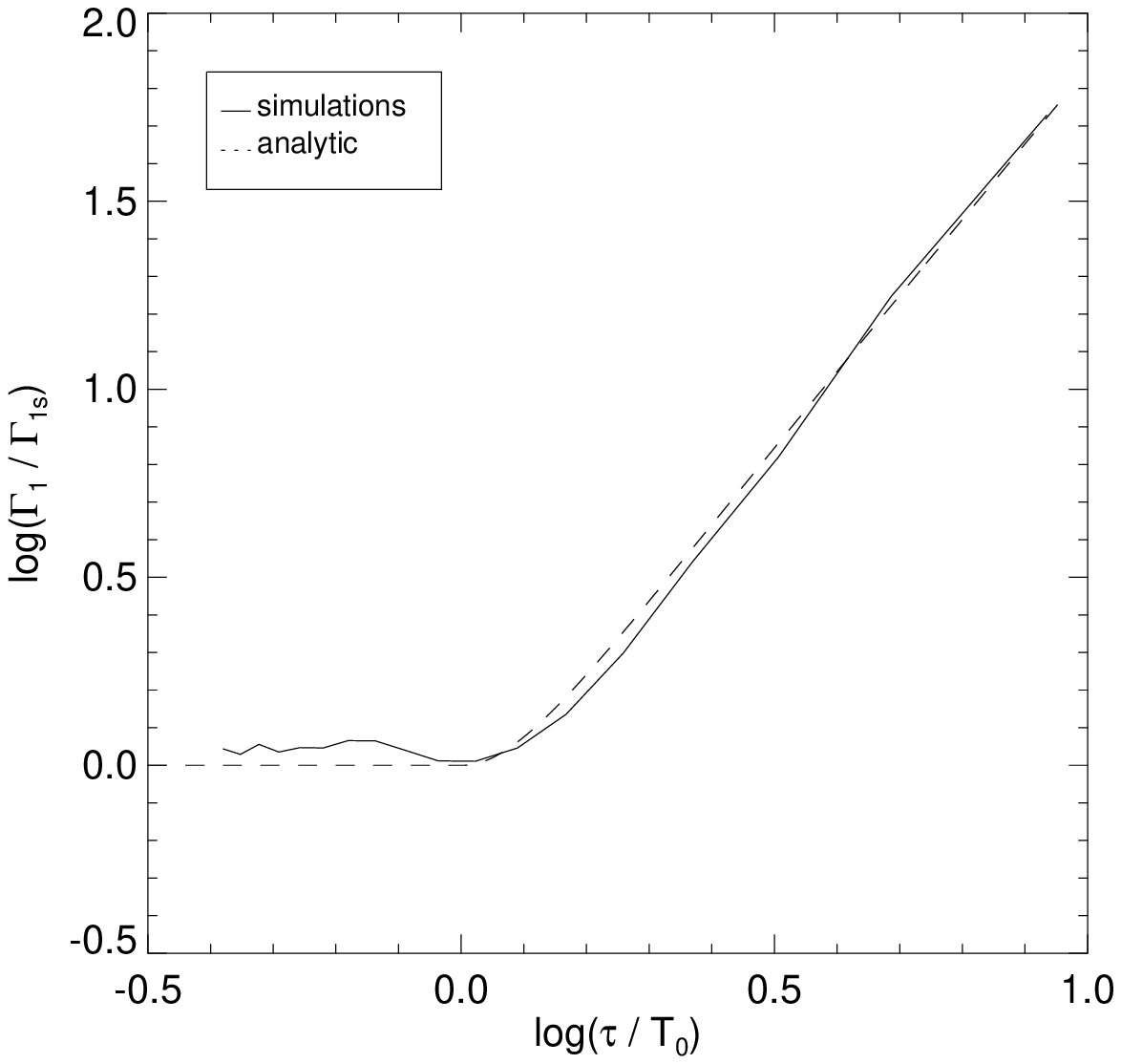}
\caption{A  plot of $\log(\Gamma_{1}/\Gamma_{1s})$ versus
$\log(\tau/T_0)$. $\Gamma_{1}$ is the value of $\vert \gamma \vert$ that is
exceeded in 1\% of the simulations. 
See caption to Figure~\ref{fig:K1} for details of the simulations.  
The solid line shows the behavior of
the simulated data, and the dashed line represents the
analytic expression from Equation \ref{eq:GammaLong}.
\label{fig:GAMMA}}
\end{figure}

Equipped with the behaviors of $V_{c_1}$ and $V_{s_1}$, we are now
in a position to write down an analytical expression for
$K_1$ ($\propto V_{c_1}^2 + V_{s_1}^2$) at all periods:
\begin{equation}
K_1 = \cases{K_{1s}  = \frac{18.42 \sigma_0^2}{n_0} & for $\tau \le T_0$ \cr 
 \frac{4 K_{1s}}{(1-\cos\left[\frac{\pi T_0}{\tau}\right])^2} & 
for $T_0 < \tau < 2T_0$ \cr
\frac{5K_{1s}}{4}\left\{\frac{4}{(1-\cos\left[\frac{\pi T_0}{\tau}\right])^2} +
\frac{1}{\sin^2\left(\frac{\pi T_0}{\tau}\right)}\right\} & for $\tau > 2T_0$ \cr} . 
\label{eq:K1long}
\end{equation}
Here, the factor of 5/4 is chosen to enforce continuity of $K_1$
for all periods.  
Since $V_c \gg V_s$ for $\tau > T_0$,
to a very good approximation, we can express Equation~\ref{eq:K1long} as
\begin{equation}
K_1 = \frac{4 K_{1s}}{(1-\cos\left[\frac{\pi T_0}{\tau}\right])^2};
{\rm \:\:\:\:for\:} \tau > T_0.
\label{eq:K1long-simple}
\end{equation}
We prefer this compact formula over the more exact formula 
(Equation~\ref{eq:K1long}).
As demonstrated by Figures \ref{fig:K1} and \ref{fig:VC}--\ref{fig:GAMMA},
the analytical expressions derived above provide excellent fits
to the simulated data. 
Thus, we have fulfilled our original objective of determining an
analytical understanding of $K_1$.  

We now turn to the importance of the phase term. As can be seen from
Figure~\ref{fig:corrls}, in the short period regime, 
the distribution of $v_c$ and $v_s$ is
cylindrically symmetric.  However, in the long-period
regime the distribution becomes highly elliptical, and the
Least Squares fit overwhelmingly prefers orbits with $\phi=\pm
90^\circ$.  An orbit sampled at a phase of $\pm 90^\circ$ has the smallest
radial acceleration and thus for a given set of measurements yields the
largest value of $K$.   

We have so far considered only the statistics of $K$, which is perfectly
reasonable when the $v_c$--$v_s$ distribution is cylindrically symmetric.
However, in the long-period regime this 
distribution is elliptical, in which case the exclusive use of the 
radial parameter $K$ is bound not to be optimal.
Thus in the long-period regime,
we need to look at both the $v_c$ and $v_s$, or equivalently
the phase and amplitude of the fitted parameters.

To this end, we define ellipses in the $v_c$--$v_s$ plane, $\epsilon_1$,
such that 1\% of simulated fitted pairs lie outside 
this ellipse.  The parameters of these ellipses are easily obtained
since we have analytical expressions for $V_{c_1}$ and $V_{s_1}$
(Equations~\ref{eq:VcLong} and \ref{eq:VsLong}). Further discussion and
exploitation of this ``amplitude-phase'' ($K$-$\phi$) analysis is postponed to 
Section \ref{sec:typeII-errors}.

\noindent{\bf Uneven and Sparse Sampling.}
We re-iterate that the above results have been derived under the 
assumption of dense, even sampling.  Here, we examine the validity
of these results in the case of random sampling, paying particular
attention to the regime where $n_0$ is small.

As discussed in Section \ref{sec:LSq}, Equations \ref{eq:V1s}--\ref{eq:K1s}
are only valid if the chosen sampling scheme preserves the independence
of $v_c$, $v_s$, and $\gamma$.  If we have a large number of samples
over a number of cycles, then even if our sampling is completely random,
we expect that the dense sampling of the sinusoidal summations will
preserve the diagonality of Equation \ref{eq:NormalEqs} (and hence
Equations \ref{eq:vc-short}--\ref{eq:gamma-short} will hold).
However, for sparse
sampling ({i.e.} small $n_0$), our choice of sampling scheme becomes
important.  In particular, if we sample the data evenly in the interval
$[-T_0/2,T_0/2]$, then Equation \ref{eq:NormalEqs} remains diagonal,
and the independence of $v_c$, $v_s$, and $\gamma$ is preserved.
In contrast, if the data are sparsely {\em and} unevenly sampled, then
the summations of $\sin(\omega t)$, $\cos(\omega t)$, and $\sin(\omega t)
\times \cos(\omega t)$ may no longer average to zero, and $v_c$,
$v_s$, and $\gamma$ will be covariant.  Physically, this corresponds to
the fact that for a small number of randomly spaced samples, the sampled
sinusoid might look like a straight line ({e.g.} if all the samples 
happen to lie at the zero-crossings of the sine-wave).
Thus, for sparse and uneven
sampling, Equations \ref{eq:V1s}--\ref{eq:K1s} do not accurately
describe the sensitivity of the Least Squares fit.

\begin{figure}[t]
\vspace{5.0 in}
\includegraphics{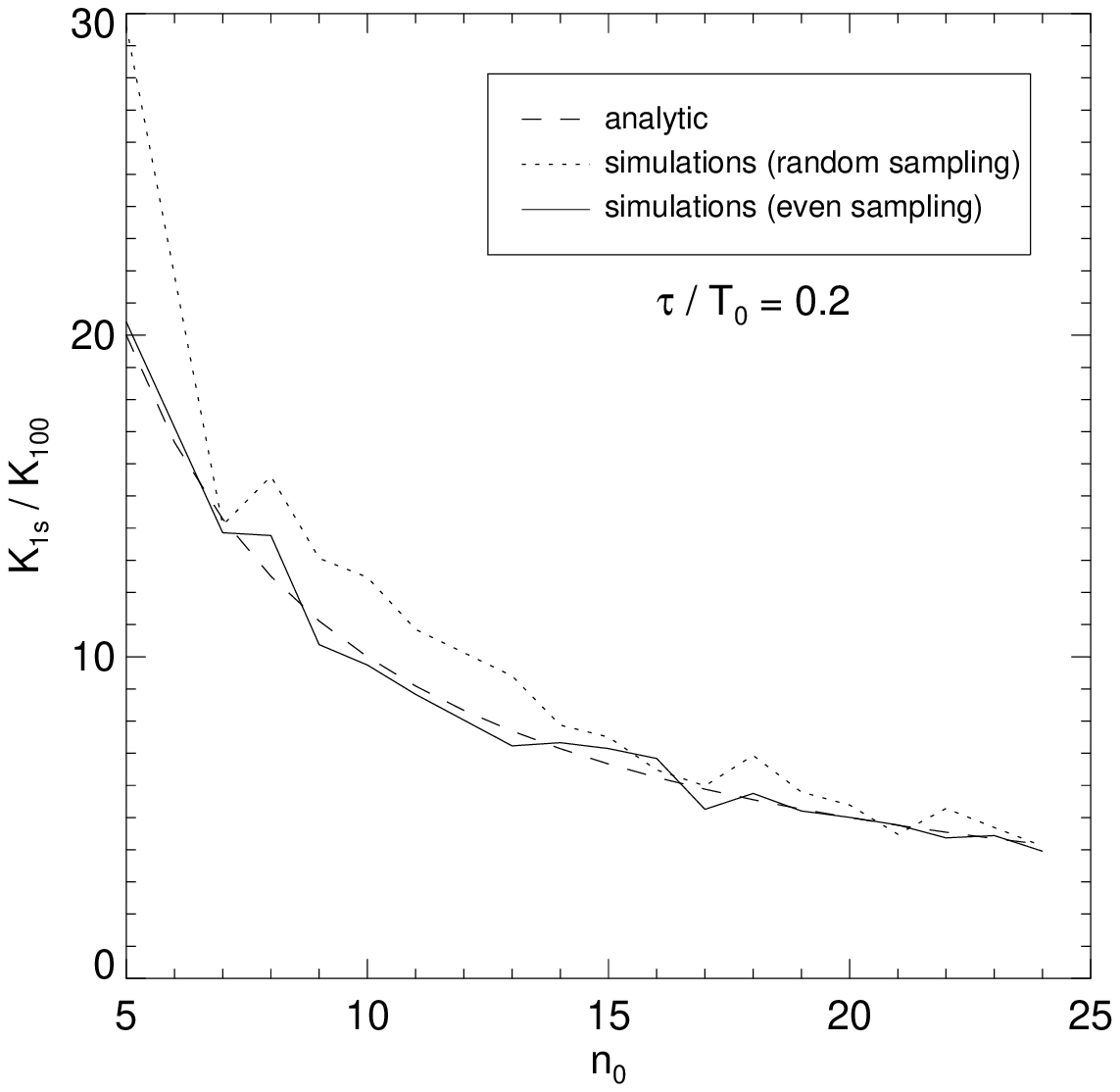}
\caption{A plot of $K_{1s}$ versus $n_0$ for evenly sampled
(solid line) and randomly sampled simulated data (dotted line).  
(For details of the simulations, see the caption to Figure
\ref{fig:K1}.)
The analytical expression given by Equation \ref{eq:K1s} is
indicated by the dashed line.  All of these curves
are normalized by the value of $K_{1s}$ for $n_0 = 100$, 
denoted by $K_{100}$.
We note that for $n_0 \lesssim 10$, the inclusion of $\gamma$ in
the model equation is important for randomly
sampled data, and that regardless of
the chosen sampling scheme, we
suffer significant losses in sensitivity.
\label{fig:K1-n0}}
\end{figure}

We verify these assertions through simulations: for $n_0=\{5,...,25\}$,
we simulate $N=1000$ data sets and do Least Squares fits to sinusoids
with $\tau=0.2\:T_0$. The sampling times are 
drawn from a distribution given by $t_j \: \epsilon \:
[\Delta t \times (j - R), \Delta t \times (j + R)]$.  
Here, $\Delta t$ denotes the sampling interval, given by $T_0 / n_0$,
and $R$ is a parameter describing the unevenness of the sampling.
$R=0$ gives an even sampling scheme, and $R=1/2$ gives 
random sampling.  The results of these simulations are shown in
Figure \ref{fig:K1-n0}.  We see that for $n_0 \lesssim 10$, one suffers 
substantial losses in sensitivity if the data are unevenly sampled.
This phenomenon was noted by Cumming et al. (1999)
\nocite{CMB99}. However, the basic reason why we see this phenomenon
is because Equation \ref{eq:K1s} is no longer valid. This discussion
re-emphasizes of the generality and robustness 
of the Least Squares approach over
the periodogram approach.

\begin{figure}
\vspace{5.0 in}
\includegraphics{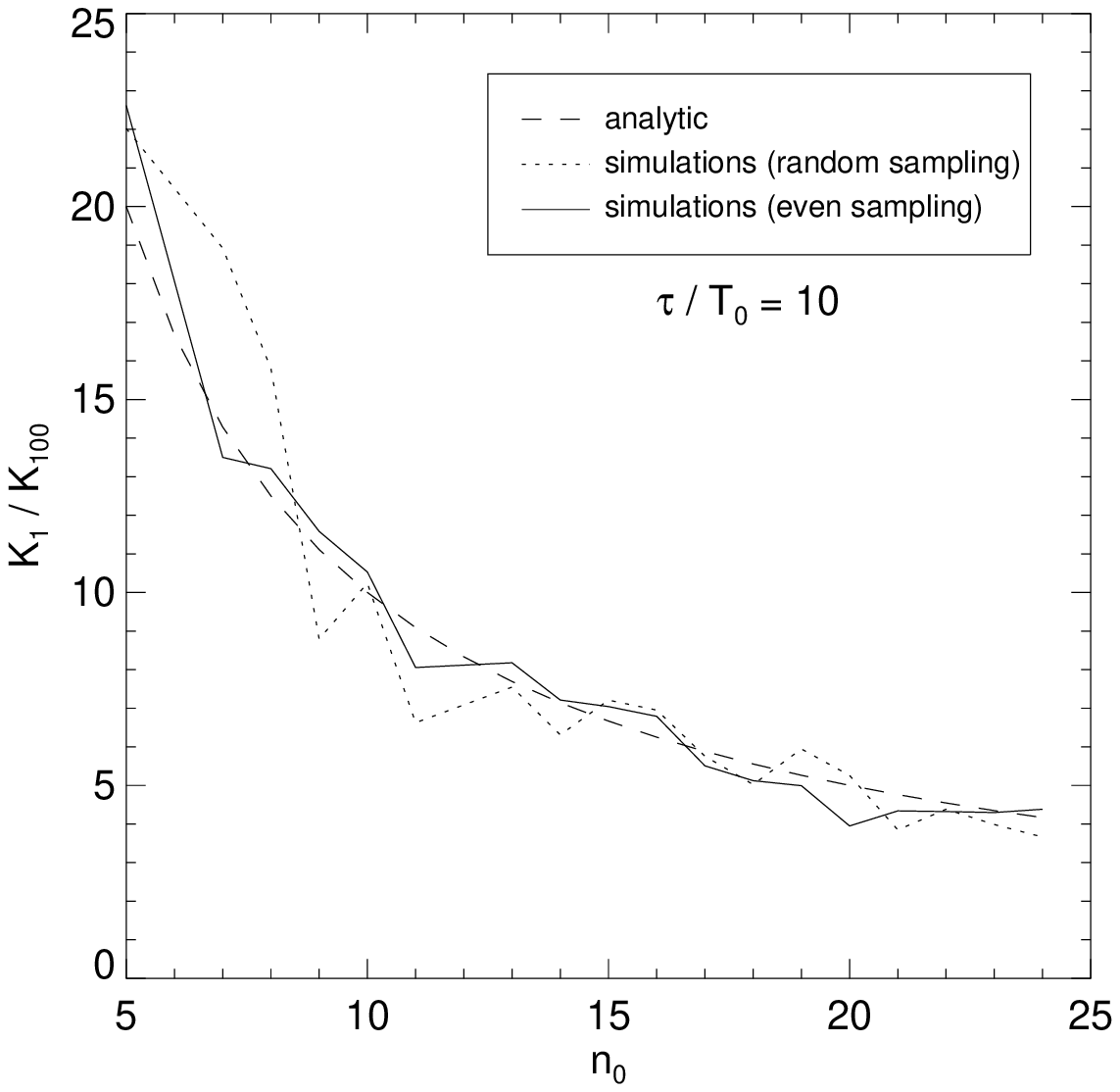}
\caption{A plot of $K_1$ versus $n_0$ for evenly sampled
(solid line) and randomly sampled simulated data (dotted line).  
(For details of the simulations, see the caption to Figure
\ref{fig:K1}.)  The analytical expression given by Equation 
\ref{eq:K1long-simple} is indicated by the dashed line.  All of these curves
are normalized by the value of $K_{1}$ for $n_0 = 100$, 
denoted by $K_{100}$.  We note that the difference between
even and random sampling is not overly important, as
we expect because the behavior of $K_1$ in the long-period regime
(Equation \ref{eq:K1long-simple}) was derived
under the assumption that $v_c$, $v_s$, and $\gamma$ are covariant.
\label{fig:K1-n0-LPR}}
\end{figure}

We now repeat the above simulations using fitted sinusoids with 
$\tau = 10 \: T_0$, to explore
the effects of sparse and uneven sampling in the long-period regime.
The results of this simulation are shown in Figure \ref{fig:K1-n0-LPR}.
From this figure we see that in the long-period regime,
the evenness of the sampling scheme is not nearly as important
as it is in the short-period regime.  This is
to be expected, because our analysis of the behavior of $K_1$ in the
long-period regime takes into account the covariance between $v_c$,
$v_s$, and $\gamma$,  and thus we expect it to be applicable 
to sparsely and unevenly sampled data.

We end this section by acknowledging that our treatment of the
statistics is not accurate in the low $n_0$ regime. At the very
least one needs to be aware of the loss of the degrees of freedom
because three parameters are obtained from the data. The correct
value of the degrees of freedom will depend on what statistic
is being estimated. This issue is not central to the main goal
of this paper and we intend to investigate this issue in a later
paper.

\section{Type II Errors}
\label{sec:typeII-errors}

So far we have computed the probability of detecting an apparent
signal  generated purely by noise.  In the language of inference, we
have discussed Type I probabilities.  Our
analysis was centered on obtaining the statistics of $K$,
the squared amplitude from the fitted parameters $v_c$ and $v_s$.  In
particular, we estimated $K_1$, the 99$^{\rm th}$ percentile of $K$ (see
Equation~\ref{eq:K1long-simple}).  However, following this analysis,
we noted that in the
long-period regime, $v_c$ and $v_s$ form an elliptical distribution,
and thus merely looking at the statistics of $K$, a radial parameter,
was not optimal.

We now consider Type II probabilities -- the probability of failing to
detect a genuine signal due to contamination by noise.
The goal of this section is to understand the statistics
of $K$ in the presence of both signal and noise.
To this end, we simulate a data set that consists of signal and noise:
\begin{equation}
v'(t_i) = \sqrt{K} \sin(\frac{2\pi t_i}{\tau} + \phi) + N(t_i),
\label{eq:BasicTII}
\end{equation}
where $\sqrt{K}$ is the amplitude of the signal,  and $N(t_i)$ is the
Gaussian noise.  We let $\phi$ be drawn from a uniform distribution in
the interval from $[0,2\pi]$, an appropriate assumption for circular
orbits.  We choose an initial signal amplitude of $\sigma_0/2$, and
then do $N = 1000$ Least Squares fits (with the same parameters as in
Section \ref{sec:typeI-errors}).

\subsection{Amplitude-Only Analysis \label{sec:amplitude-only}}

\begin{figure}
\vspace{4.5 in}
\includegraphics{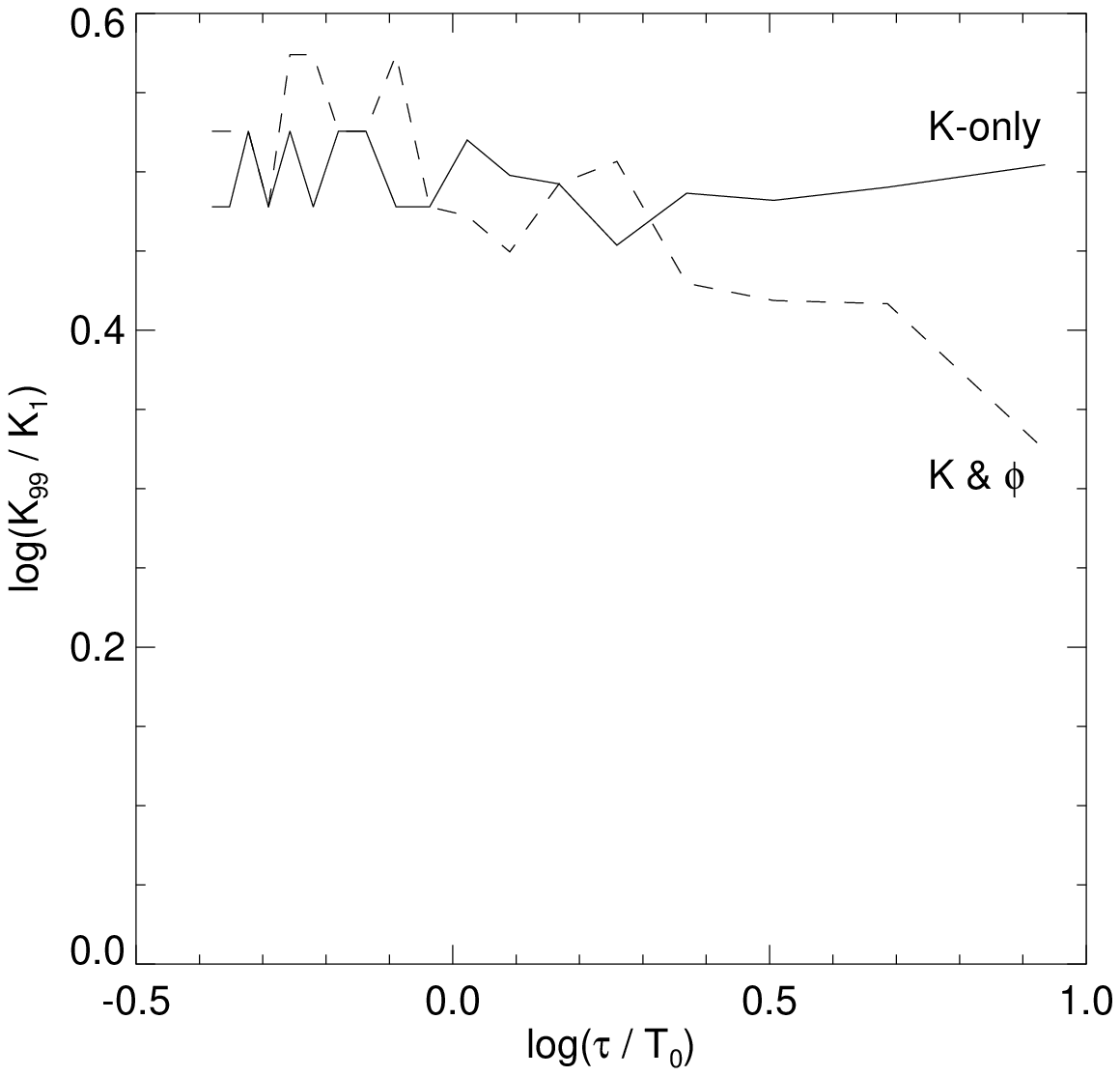}
\caption{A plot of $\log(K_{99}/K_1)$
versus $\log(\tau/T_0)$ as determined
in the amplitude-only and $K$-$\phi$ cases.
For the amplitude-only
case (solid line), $K_{99}$ represents the squared
signal amplitude such that 99\% of the simulated data yield fitted
$K$'s greater than $K_1$.  For the $K$-$\phi$ analysis, $K_{99}$
is the necessary amplitude such that 99\% of fitted $K$ and $\phi$, or
equivalently 
$\{v_c,v_s\}$, lie outside of $\epsilon_1$ (the $\epsilon_1$
ellipse contains 99\% of fits to noise-only data; for further
details see the discussion towards the end of 
Section \ref{sec:typeI-errors}).  The 99$^{\rm th}$ percentile of the 
slope analysis (discussed in Section \ref{sec:linear}) is greater
than the amplitude-only $K_{99}$ for all sampled periods, 
and thus is not shown here.  At each period, we carried
out 10,000 simulations; the phase of the signal was assumed to be
randomly and evenly distributed over the range $[0,2\pi]$.  
\label{fig:typeII-errors}} 
\end{figure}

Following the path used for Type I errors, we will first base
our analysis on $K$, i.e. we will ignore the issue of the elliptical
nature of the $v_c$--$v_s$ distribution. At each period,
we determine how many of
the fitted amplitudes lie below $K_1$.  If it is less
than $0.01N$, then we have found the value of $K$ such that 99\% of the
fitted amplitudes lie above $K_1$.  We call this $K_{99}$.  If,
however, more than 1\% of the fitted amplitudes lie below $K_1$, we
increment the signal amplitude by $K^{0.6} / 20$ until we find $K_{99}$.  
A plot of $K_{99}/K_1$ versus period is shown in Figure
\ref{fig:typeII-errors}.

\begin{figure}
\vspace{4.5 in}
\includegraphics{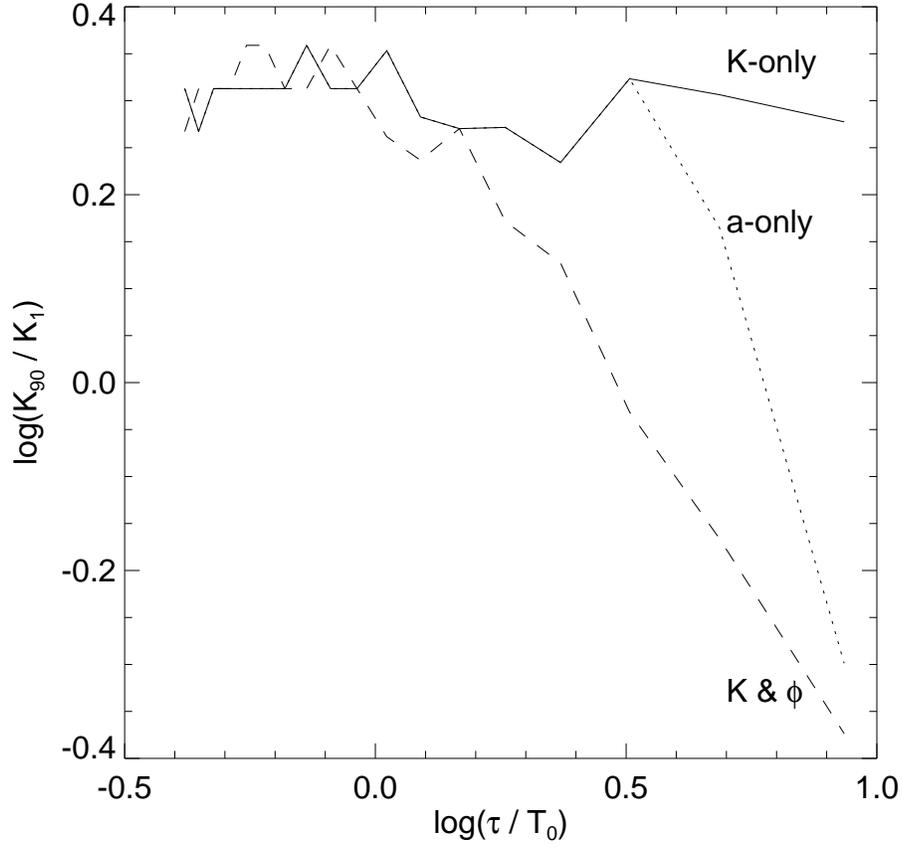}
\caption{A plot of $\log(K_{90}/K_1)$ versus $\log(\tau/T_0)$ as
determined in the amplitude-only (solid line),
$K$-$\phi$ (dashed line), and slope-only (dotted line) cases. 
At sampled periods where the amplitude-only $K_{90}$ is
less than the slope-only $K_{90}$, only the former is shown.
$K_{90}$ represents the squared signal amplitude such that 90\% 
of the simulated data yield
fitted coefficients greater than the type I sensitivity limits 
(at that period), given by $K_1$, $\epsilon_1$, and $A_1$ for the
$K$-only, $K$-$\phi$, and $a$-only analyses, respectively. 
See the caption to Figure~\ref{fig:typeII-errors} for further details.
\label{fig:K90}}
\end{figure}

\begin{figure}
\vspace{4.5 in}
\includegraphics{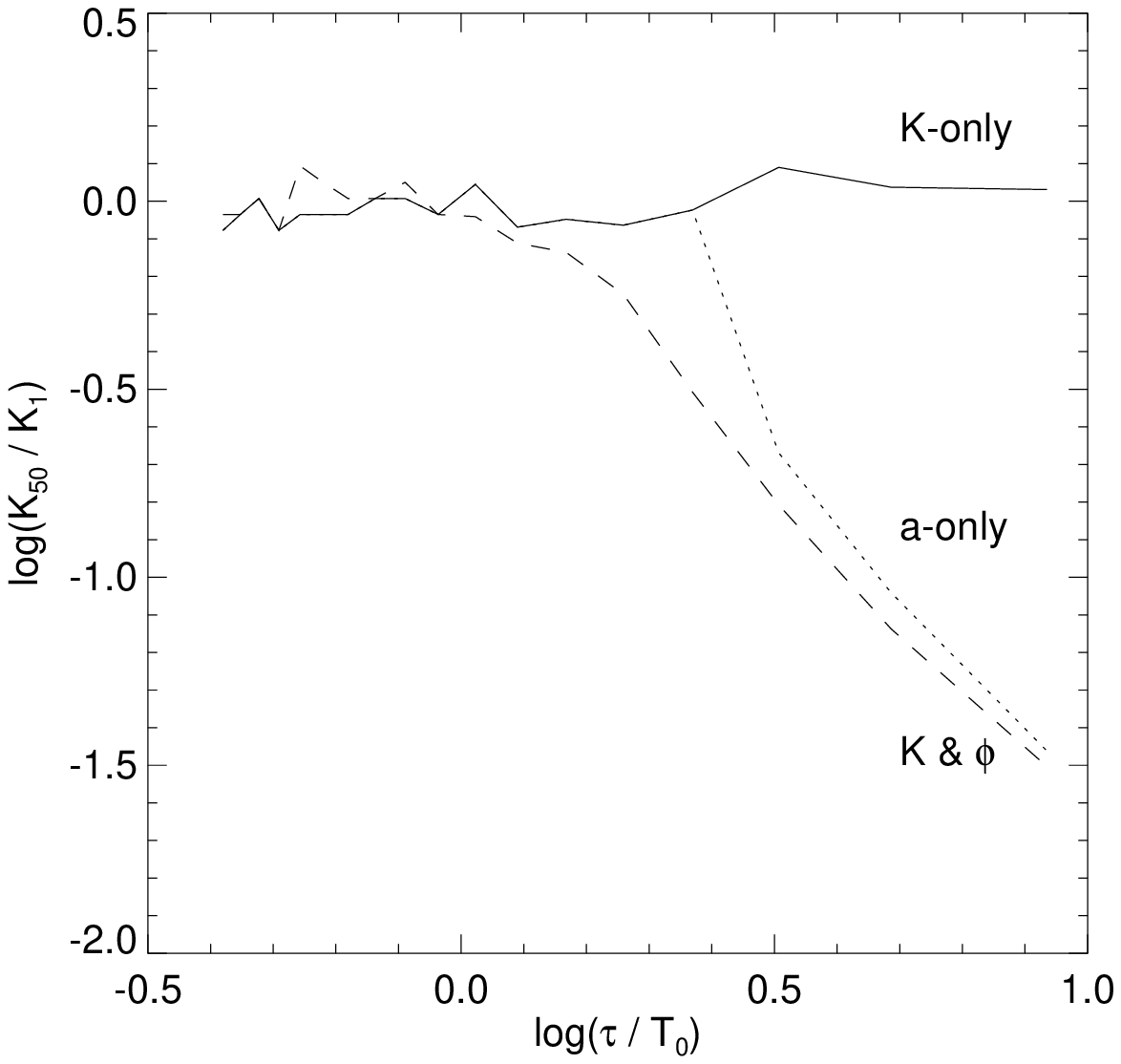}
\caption{A plot of $\log(K_{50}/K_1)$ 
versus $\log(\tau/T_0)$ as determined
in the amplitude-only case (solid line), the $K$-$\phi$ case
(dashed line), and the slope-only case (dotted line).  
At sampled periods where the amplitude-only $K_{50}$ is
less than the slope-only $K_{50}$, only the former is shown.
$K_{50}$ represents the squared
signal amplitude such that 50\% of the simulated data yield
fitted coefficients greater than the type I sensitivity limits 
(at that period),  given by $K_1$, $\epsilon_1$, and $A_1$ for the
$K$-only, $K$-$\phi$, and $a$-only analyses, respectively. 
See the caption of Figure \ref{fig:typeII-errors} for further details.
\label{fig:K50}} 
\end{figure}

We note that our choice of 99\% confidence for $K$ is arbitrary and
also very conservative. Most observers would be keen to make a
discovery rather than set stringent upper limits. Thus one may wish to
consider $K_{50}$ (or $K_{90}$), which gives the signal
amplitude necessary to be detected 50\% (or 90\%) of the time.  Plots
of $K_{90}$ and $K_{50}$ versus period are shown in
Figure~\ref{fig:K90} and Figure~\ref{fig:K50}, respectively.

\subsection{Amplitude-Phase Analysis \label{sec:amplitude-phase}}

\begin{figure}
\vspace{5.2in}
\includegraphics{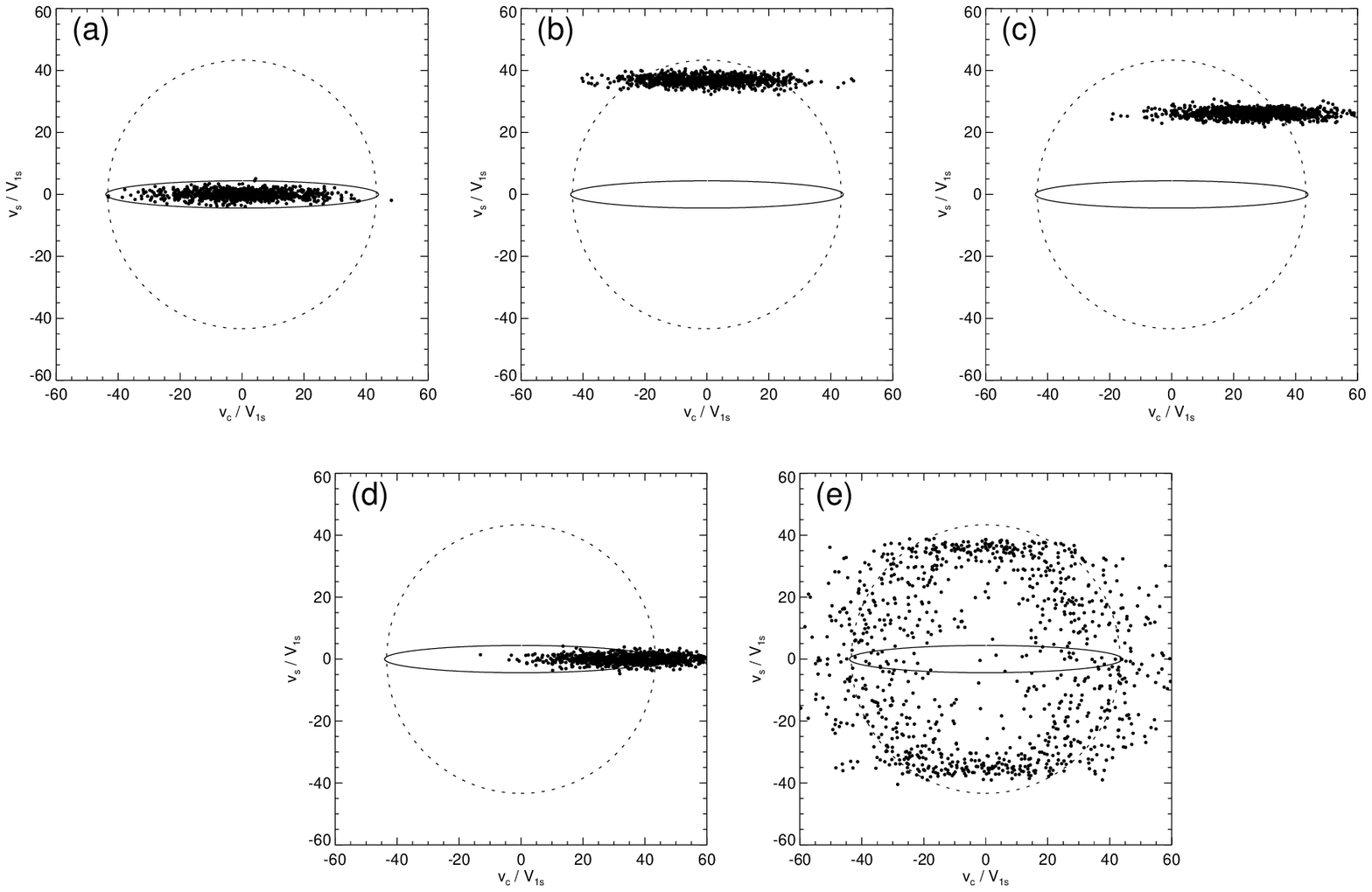}
\caption{Distribution of the fitted parameters, $\{v_c,v_s\}$ for
$\log(\tau_0/T_0)=1.0.$
A total of $N=1000$ simulations were performed for five different
assumptions.
(a) No signal is
present.  (b)--(e) show data sets containing signals
with an amplitude of $\sqrt{K_1}$, and $\phi=0^{\circ}$,
$45^{\circ}$, $90^{\circ}$, and random phase, respectively.  
$\epsilon_1$ and the circle of radius $K_1$ are also
plotted in (a)--(e).  (a) shows that 99\% of the fitted data
points lie within $\epsilon_1$.  (b)--(e) demonstrate the
importance of phase in the determination of Type II errors:
for signals with phases different than $\pm 90^\circ$,
the fitted $\{v_c,v_s\}$ have a significant component along the
$v_s$-axis, causing them to lie outside of $\epsilon_1$.
Thus, the fraction of signals detected by the $K$-$\phi$
analysis is (100\%, 100\%, 33\%, 95\%) for cases (b)--(e), while
the corresponding fractions for the amplitude-only analysis are 
(25\%, 35\%, 33\%, 34\%).
\label{fig:e1-k1-signals}}
\end{figure}

In Section \ref{sec:typeI-errors}, we showed that in the
long-period regime, the $v_c$--$v_s$ distribution is elliptical. In view
of this, the phase of the signal in Equation~\ref{eq:BasicTII} is
critical. This point is best understood in the results of the
simulations displayed in  Figure~\ref{fig:e1-k1-signals} for five
cases:  (a) no signal, (b)--(d), a signal with amplitude $\sqrt{K_1}$
(at that period) and $\phi=0^\circ, 45^\circ, 90^\circ$, respectively, and
(e) a signal with amplitude of $\sqrt{K_1}$ and the phase being
randomly (for each simulation)
chosen from the range $[0,2\pi]$ (uniform probability density
function).

In Figure~\ref{fig:e1-k1-signals}, the amplitude-only analysis would
consider all points that lie inside the circle of radius $K_1$ to be
indistinguishable from those produced by noise.  However, in the
framework of the $K$-$\phi$ analysis, we would consider all points
within the ellipse to be indistinguishable from those produced from
noise.  The superiority of the $K$-$\phi$ analysis is evident
given that the size of the ellipse is smaller than that of the circle.
This superiority is reflected numerically as follows:
the fraction of simulations in which the signal is reliably recovered
by the $K$-$\phi$ analysis is (100\%, 100\%, 33\%, 95\%) 
for cases (b--e) whereas for the
amplitude-only analysis the corresponding fractions
are (25\%, 35\%, 33\%, 34\%).

A comparison of Figures~\ref{fig:typeII-errors}--\ref{fig:K50}
shows the increasing gain of the $K$-$\phi$
analysis as the tolerance for committing type II error is increased.
When the Type II probability is set to 99\%, the $K$-$\phi$ analysis
lowers the $K$ threshold by 20\% whereas if we are willing to accept a
50\% type II error probability then the $K$ threshold is lowered by
nearly a factor of 30!

\subsection{Linear Analysis of Long Term Trends}
\label{sec:linear}

Above we illustrated the superiority of an amplitude-phase analysis
over an amplitude-only analysis.
However, some previous authors ({\cite{WALKER+95}; \cite{CMB99})
have analyzed long-term trends in the data ({i.e.} apparent signals
with $\tau \gg T_0$) by examining the slope of the best-fit straight
line, denoted by $a$.  This best-fit line is subtracted from
the data, and an amplitude-only analysis is subsequently performed
on the residuals. Here, we discuss this ``slope analysis'' in the
context
of the results of Sections \ref{sec:amplitude-only} and
\ref{sec:amplitude-phase}.

To determine the sensitivity of slope analysis, we simulate 1000
data sets with Gaussian noise of zero mean and $\sigma_0=3 {\rm \:m
\: s^{-1}}$, sampled at one month intervals for
$T_0=12$ years.  For each simulated data set, we perform
a Least Squares fit to a straight line, given by $at+b$.
The 99$^{\rm th}$ percentile slope, denoted by $A_1$, is
determined  by finding the value of $\vert a \vert$
such that 99\% of simulated data sets yield $\vert a \vert < A_1$.
Next, we determine the type II errors, by injecting a sinusoidal
signal into the simulated data (Equation \ref{eq:BasicTII})
and finding the necessary squared signal amplitude ($K$)
such that the signal can be reliably recovered.
This analysis is the same as that carried out in Section
\ref{sec:amplitude-only} except that instead of using the
amplitude of fitted sinusoids, we use the slope of fitted
straight lines to set confidence limits.  Thus, at each sampled
period, we can distinguish a real detection from one produced by noise
if the fitted $\vert a \vert$ is greater than $A_1$.

The results of these simulations are shown in Figures
\ref{fig:typeII-errors}--\ref{fig:K50}.  At sampled periods
where the amplitude-only analysis is more sensitive than the slope
analysis, only the former is shown.  We note that
although the slope analysis may be more sensitive than the
amplitude-only analysis (in the long-period regime),
the amplitude-phase analysis still yields a significant improvement.
Moreover, the discrepancy between the amplitude-phase and slope analyses
is largest for $\tau \lesssim 3 \: T_0$.
We examine this discrepancy further by performing
an analysis analogous to that shown in Figure \ref{fig:e1-k1-signals}e.
In particular, for fitted periods of $\tau=2\: T_0$ and $\tau=10\: T_0$,
we inject simulated data sets consisting of sinusoids of random phase
and
$K=K_1$ plus Gaussian noise.    The percentage of signals recovered by
the amplitude-only, amplitude-phase, and slope analyses (respectively)
is (57\%, 87\%, 59\%) for $\tau=2\: T_0$, and (34\%, 95\% 91\%) for
$\tau=10\: T_0$.

We can understand the relative sensitivities of the $K$-$\phi$
and slope analyses
by noting that the slope analysis is an explicitly linear technique,
and thus it throws away any information that is
contained in the curvature of the sine or cosine components of
the fitted sinusoid. The $K$-$\phi$ analysis, in contrast,
utilizes all of the information contained in the linear {\em and}
curvature components of the fitted sinusoid.  These curvature components
will tend to zero for $\tau \gg T_0$, and thus in this regime the
sensitivity of the slope analysis will approach that of amplitude-phase
analysis.  However, for $\tau \sim 2-3 \: T_0$, these curvature terms
are significant, and the $K$-$\phi$ analysis yields a substantial
improvement over a slope analysis.

The above discussion (in our view) re-emphasizes the generality of
the Least Squares approach. While Walker et el. (1995) and
Cumming et al. (1999)\nocite{CMB99} introduce
a modified periodogram to account for a slope in the data,
the Least Squares approach can be applied without modification,
as it has been in \S\ref{sec:amplitude-phase}.

\section{Conclusions}
\label{sec:conclusions}
We have developed an analytical understanding of the sensitivity 
of the radial velocity technique of planet detection in the regime
where the orbital period is longer than the total baseline of the
observations.  We also examined the sensitivity in the 
short-period regime, paying particular attention to the case of
sparsely sampled data.  Moreover, we
have illustrated the benefits of Least Squares fitting over the
equivalent, but more complicated technique of periodogram analysis;
while the periodogram must be modified to deal with
long-period signals, or with sparsely sampled data, the Least Squares
approach can be applied in its basic form.

We have also discovered the potentially exciting  new result that in the 
long-period regime one obtains additional information from the phase. 
Analyses of the RV data to date ({e.g.} \cite{WALKER+95}; \cite{NA98}; 
\cite{CMB99}) have been based either 
on amplitude-only analyses, or on amplitude analyses supplemented
by slope analyses. However, as dramatically demonstrated by
Figures~\ref{fig:K90} and \ref{fig:K50},  $K$-$\phi$ analysis
has significant advantages over these previous analyses, especially
in the interesting regime where $\tau \sim 2-3 \: T_0$. 

Thus, we propose using a confidence test based on amplitude and phase,
characterized by ellipses in the $v_c$--$v_s$ plane, $\epsilon_1$.
The analytical expressions we have developed for $V_{c_1}$ and
$V_{s_1}$ provide the analytical behavior of $\epsilon_1$ as a 
function of period.  For real RV data, like those mentioned above,
we suggest that the fitted $v_c$ and $v_s$ (at each
period) be compared to the $\epsilon_1$ ellipses. Points lying
outside these ellipses would signify the detection of periodic signals.

We have applied this $K$-$\phi$ analysis technique to the
RV data of Walker et al. (1995).  While Nelson \& Angel (1998)
examined this data with an amplitude-only analysis and reported 
several marginal detections, our technique yielded several clear
detections of periodic signals.  Although we cannot say whether
these periodicities represent planetary signals (as opposed to
stellar cycles or periodic systematic errors),
it is clear that the use of amplitude and phase allows 
previously unknown periodicities to be detected.  In the future,
we hope to apply this $K$-$\phi$ analysis to more
comprehensive RV surveys, and to detect long-period companions.

\end{document}